\documentstyle[12pt]{article}

\def\({[}
\def\){]}
\def\CP{{\em CP}~}
\def\cp{{\em CP}}
\def\sa{\sin 2\alpha}
\def\sb{\sin 2\beta}
\def\GeV{{\rm GeV}}
\def\MeV{{\rm MeV}}
\def\B{{\rm B}}
\def\K{{\rm K}}
\def\roughly#1{\raise.3ex\hbox{$#1$\kern-.75em\lower1ex\hbox{$\sim$}}}
\def\be{\begin{equation}}
\def\ee{\end{equation}}
\def\bea{\begin{eqnarray}}
\def\eea{\end{eqnarray}}

\begin{document}
\begin{titlepage}
\begin{center}
          \hfill    WIS-92/52/Jun-PH \\
          \hfill    LBL-32563 \\
          \hfill    hep-ph/9207225 \\
          \hfill    July 1992 \\
          \hfill    (T/E)
\vskip .25in

{\large \bf
Testing the Standard Model and
Schemes for Quark Mass Matrices with
\CP Asymmetries in B Decays}

\vskip .25in

Yosef Nir\footnote{Incumbent of The Ruth E. Recu Career
Development Chair. Supported in part by the Israel Commission
for Basic Research, United States--Israel Binational
Science Foundation and the Minerva Foundation.
E-mail: ftnir@weizmann.weizmann.ac.il}\\[.1in]

{\em Physics Department, Weizmann Institute of Science,
 Rehovot 76100, Israel }

\vskip 9pt
Uri Sarid\footnote{E-mail: sarid\%theorm.hepnet@lbl.gov}\\[.1in]

{\em Physics Department, Weizmann Institute of Science,
 Rehovot 76100, Israel\\
           {\rm and}\\
      Theoretical Physics Group\footnote{Permanent address}\\
      Lawrence Berkeley Laboratory\\
     1 Cyclotron Road\\
     Berkeley, California 94720}
\end{center}

\vskip .2in

\begin{abstract}

The values of $\sa$ and $\sb$, where $\alpha$ and $\beta$ are angles
of the unitarity triangle, will be readily measured in a B factory (and maybe
also in hadron colliders).
We study the standard model constraints in the $\sa-\sb$ plane. We
use the results from recent analyses of $f_B$ and $\tau_b|V_{cb}|^2$
which take into account heavy quark symmetry considerations.
We find $\sb\geq0.15$ and most likely $\sb\,\roughly{>}\,0.6$, and emphasize
the
strong correlations between $\sa$ and $\sb$.
Various schemes for quark mass matrices allow much smaller areas in the
$\sa-\sb$ plane. We study the schemes of Fritzsch, of Dimopoulos, Hall and
Raby, and of Giudice, as well as the ``symmetric CKM'' idea, and show how \CP
asymmetries in B decays will crucially test each of these schemes.

\end{abstract}
\end{titlepage}

\renewcommand{\thepage}{\roman{page}}
\setcounter{page}{2}
\mbox{ }

\vskip 1in

\begin{center}
{\bf Disclaimer}
\end{center}

\vskip .2in

\begin{scriptsize}
\begin{quotation}
This document was prepared as an account of work sponsored by the United
States Government.  Neither the United States Government nor any agency
thereof, nor The Regents of the University of California, nor any of their
employees, makes any warranty, express or implied, or assumes any legal
liability or responsibility for the accuracy, completeness, or usefulness
of any information, apparatus, product, or process disclosed, or represents
that its use would not infringe privately owned rights.  Reference herein
to any specific commercial products process, or service by its trade name,
trademark, manufacturer, or otherwise, does not necessarily constitute or
imply its endorsement, recommendation, or favoring by the United States
Government or any agency thereof, or The Regents of the University of
California.  The views and opinions of authors expressed herein do not
necessarily state or reflect those of the United States Government or any
agency thereof of The Regents of the University of California and shall
not be used for advertising or product endorsement purposes.
\end{quotation}
\end{scriptsize}

\vskip 2in

\begin{center}
\begin{small}
{\it Lawrence Berkeley Laboratory is an equal opportunity employer.}
\end{small}
\end{center}

\newpage
\renewcommand{\thepage}{\arabic{page}}
\setcounter{page}{1}

\CP asymmetries in neutral $\B$ decays will provide a unique
way to measure the CKM parameters. In a high-luminosity
$e^+e^-$ collider running at the energy of the $\Upsilon(4S)$
resonance (a ``$\B$ factory''), two of the three angles of the
unitarity triangle (see Fig.~1) will be readily measured \cite{carsan}:
the \CP asymmetry
in {\it e.g.} $\B\rightarrow\pi^+\pi^-$ will determine $\sa$, while
that in {\it e.g.} $\B\rightarrow\psi \K_S$ will determine
$\sb$. It may also be possible to measure $\sb$ in a hadron collider, but $\sa$
would be difficult due to the large background (see, {\it e.g.}, \cite{nirq}).
The experimental measurements are expected
to be highly accurate and the theoretical calculations are,
to a large extent, free of hadronic uncertainties.
Furthermore, \CP asymmetries in neutral $\B$ decays
are a powerful probe into possible sources of \CP violation
beyond the standard model (SM).
The richness of available $\B$ decay modes would allow one to determine
detailed features of the new sources of \CP violation if the
SM predictions are not borne out.
In this work, we refer to both aspects of \CP asymmetries in $\B$ decays,
namely the determination of the CKM parameters within the SM,
and the testing of extensions of the SM,
with a special emphasis on the information that can be extracted
by measuring two angles of the unitarity triangle rather than, say,
$\sb$ alone.
\par
 In the first part of this work, we investigate in detail the
SM predictions for $\sa$ and $\sb$.
In particular, we study the correlation between the two quantities
and present our results in the $\sa-\sb$ plane.
We update previous analyses with  emphasis on recent
theoretical developments which involve the heavy quark symmetry.
\par
 In the second part of this work, we
 show how various schemes for quark mass
matrices can be tested through their predictions for $\sa$ and
$\sb$. We analyze the Dimopoulos-Hall-Raby (DHR) scheme \cite{dhr},
the Giudice scheme \cite{giud}, the Fritzsch scheme \cite{frit},
and the idea that the CKM matrix is symmetric in the absolute values
of its entries \cite{brpr} (including the two-angle
parametrization of Kielanowski \cite{kiel}).
Each of these schemes allows a range for the asymmetries which is
much smaller than in the SM and thus may be clearly
excluded when the asymmetries are measured.
\par
Various bounds on the CKM parameters are usually presented as constraints
on the form of the unitarity triangle (for a review see
\cite{nir,fran,nirq}
and references therein). However, the quantities directly
measurable via \CP violation in a $\B$-factory are $\sa$ and $\sb$,
so we will present our constraints in terms of these observables.
The time-dependent \CP asymmetry in the decay of
a $\B$ or $\bar \B$ into some final \cp-eigenstate $f$ is given by
\be
{\Gamma(\B^0(t)\rightarrow f)-\Gamma(\bar \B^0(t)\rightarrow f)
\over \Gamma(\B^0(t)\rightarrow f) + \Gamma(\bar \B^0(t)\rightarrow f)}
=-{\rm Im}\lambda(f)\,\sin\Delta M\,t\,,\label{abimlam}
\ee
where $\Delta M\equiv M(\B_{\rm Heavy})-M(\B_{\rm Light})$,
$\B^0(t)$ ($\bar \B^0(t)$) is a state which starts out as the flavor
eigenstate $\B^0$ ($\bar \B^0$) at a time $t=0$, and $\lambda(f)$ is
a complex number with (almost exactly) unit magnitude. Then, within
the SM (and in all schemes considered in this work),
\be
{\rm Im}\lambda(\pi^+\pi^-)=\sa,\ \ \
{\rm Im}\lambda(\psi K_S)=\sb \label{abab}
\ee
(where we took into account the fact that $\psi K_S$ is a \cp-odd state).
Thus, our figures in the $\sa-\sb$ plane simply present the allowed range
in the ${\rm Im}\lambda(\pi^+\pi^-)-{\rm Im}\lambda(\psi K_S)$ plane.
This gives an important advantage to our method: the presentation in the
${\rm Im}\lambda(\pi^+\pi^-)-{\rm Im}\lambda(\psi K_S)$ plane
allows a direct comparison
of the SM predictions (or the experimental results)
with models of new physics where the asymmetries are not related to
angles of the unitarity triangle.
\par
We use the following relations to transform from the
$(\rho,\eta)$ coordinates of the free vertex $A$ of the unitarity
triangle to $(\sa,\sb)$:
\bea
\sa&=&{2\eta\(\eta^2+\rho(\rho-1)\)\over
\(\eta^2+(1-\rho)^2\)\(\eta^2+\rho^2\)},\nonumber \\
\sb&=&{2\eta(1-\rho)\over\eta^2+(1-\rho)^2}.\label{aba}
\eea
Note that these coordinate transformations are highly nonlinear;
hence the predictions in the $\sa-\sb$ plane will be very different
from the more familiar constraints in the $\rho-\eta$ plane. Furthermore,
since (\ref{aba}) are not (uniquely) invertible,
we may not simply map the regions
in the $\rho-\eta$ plane allowed by each of the various constraints into
corresponding regions in the $\sa-\sb$ plane, and then assume that the
overlap in the latter is allowed.
To see this, note that a single point in the
overlap region in the $\sa-\sb$ plane may correspond to two different
points in the $\rho-\eta$ plane. If each of these two points is allowed
by one constraint but forbidden by the other, then the original point in
the $\sa-\sb$ plane is in fact forbidden though it is in the overlap of
two regions allowed by the individual constraints.
We therefore form the overlap in the $\rho-\eta$ plane
first, and then map this overall-allowed region into $\sa-\sb$
coordinates. Finally, even in the $\rho-\eta$ plane the overlap
of two allowed regions may not all be allowed: a given point in the
overlap may meet the various constraints only by using different values
of some parameter which enters into both constraints. But this
correlation is unimportant in practice, since the uncertainties in
the parameters which enter into more than one constraint
never dominate both constraints.
\par
 We now analyze the SM predictions for $\sa$
and $\sb$, updating previous analyses of constraints on the CKM
parameters. The most significant update is in the constraint
from $\B-\bar \B$ mixing, which determines the length of one side
of the unitarity triangle:
\be
(1-\rho)^2+\eta^2 = {(1.3\times10^7\,\GeV) \,x_d \over
(B_Bf_B^2)y_t f_2(y_t)(\tau_B |V_{cb}|^2)|V_{cd}|^2\eta_B}\label{abacons}
\ee
where $\eta_B=0.85$ is a QCD correction, $y_t=(m_t/M_W)^2$ and
$f_2(x)=1-{3\over4}x(1+x)(1-x)^{-2}\(1+2x(1-x^2)^{-1}\ln(x)\)$.
Recently, both lattice and QCD sum-rule calculations
of the $f_B$ decay constant were made which rely on
heavy quark symmetry considerations. Results from
the two techniques now converge to a consistent range
and, we believe, should be preferred over previous,
more model-dependent, calculations. We use the result of ref.
\cite{mnfb} from QCD sum-rules,
which is consistent with lattice calculations (see \cite{agjss}
and references therein),
\be
f_B=190\pm50\,\MeV.\label{abb}
\ee
Since the $B_B$ factor is expected to be close
to unity, we simply take $B_B=1$ and neglect the uncertainty in $B_B$
relative to that in $f_B$ (or, equivalently, absorb it
into the uncertainty in (\ref{abb})).
Heavy quark symmetry considerations have also been applied to
find the combination $|V_{cb}|^2\tau_B$. We again believe that
the new results, in which only the corrections to the heavy quark limit
are model-dependent, should replace previous calculations which
were completely model dependent. We take the analysis of ref.
\cite{mnvcb} with updated input data \cite{mnpc}:
\be
|V_{cb}|\ (\tau_b/1.3\rm\,ps)^{1/2}=0.040\pm0.005\,.\label{abc}
\ee
For the mixing parameter $x_d$, we use \cite{dani}
\be
x_d=0.67\pm0.11\,.\label{abd}
\ee
Finally, we use $|V_{cd}|=|V_{us}|=0.221\pm0.002$.
\par
Our second constraint comes from the endpoint of the lepton
spectrum in charmless semileptonic $\B$ decays. We adopt the range quoted
by the Particle Data Group \cite{pdg}:
\be
|V_{ub}/V_{cb}|=0.10\pm0.03\,.\label{abe}
\ee
This determines the length of the other side of the unitarity triangle:
\be
\rho^2+\eta^2=\left|{V_{ub}\over V_{cb}V_{cd}}\right|^2
\,.\label{abbcons}
\ee
\par
The third constraint comes from the \cp-violating $\epsilon$ parameter in
the $\K^0$ system:
\be
\rho=\left\(1+{(\eta_3 f_3(y_c,y_t)-\eta_1)y_c
  \over \eta_2 y_t f_2(y_t) |V_{cb}|^2}\right\)
 - {1\over\eta} \left\(2.5\times10^{-5}|\epsilon|
  \over \eta_2 y_t f_2(y_t) |V_{cb}|^4 B_K |V_{cd}|^2\right\)\label{abeps}
\ee
where $\eta_1=0.7$, $\eta_2=0.6$ and $\eta_3=0.4$ are QCD corrections
\cite{dfp},
$y_c=(m_c/M_W)^2$ and
$f_3(x,y)=\ln(y/x)-{3\over4}y(1-y)^{-1}\(1+y(1-y)^{-1}\ln(y)\)$.
The uncertainties here lie in the value of the $B_K$ parameter,
estimated to be
\be
B_K=2/3\pm1/3,\label{abf}
\ee
and in the range for $|V_{cb}|$. Using~\cite{pdg}
$\tau_B=1.29\pm0.05\rm\,ps$, we deduce from (\ref{abc}):
\be
|V_{cb}|=0.040\pm0.007.\label{abg}
\ee
We further use $|\epsilon|=(2.26\pm0.02)\times10^{-3}$ and \cite{gassleut}
$m_c(m_c)=1.27\pm0.05\,\GeV$.
\par
Since the $x_d$ and $\epsilon$ constraints depend on $m_t$,
we have carried out our analysis for various $m_t$ values
within the range $90\,\GeV\leq m_t\leq185\,\GeV$.
We present our results in Fig.~2 in two ways. First,
the thin black curves encompass all values of $(\sa,\sb)$ which satisfy
all three
constraints using values of the input parameters within their $1-\sigma$
ranges (or within the
theoretically favored ranges for the parameters $B_K$ and $f_B$).
That is, the SM can
accommodate a $\B$-factory result anywhere within these curves without
stretching any input parameter beyond its $1-\sigma$ range. We will refer
to these regions as the ``allowed'' areas of the SM. (A somewhat similar plot
of $\sa$ versus $\rho$ appears in \cite{agjss}). Second (and similarly to
\cite{harr}), in
order to get a sense of the expected value of $(\sa,\sb)$ given our
current knowledge of the various input parameters, we generated
numerous sample values for
these parameters based on a Gaussian distribution for $|V_{cd}|$,
$\tau_B |V_{cb}|^2$, $|V_{ub}/V_{cb}|$, $\tau_B$, $x_d$, $m_c$ and
$|\epsilon|$, and a uniform distribution ($=0$ outside of the
``$1-\sigma$'' range) for $f_B$.
For each sample set we used the constraints (\ref{abacons}) and (\ref{abbcons})
to determine $\rho$ and $\eta$, and then rejected those sets
which did not meet the constraint (\ref{abeps}) for $1/3 \le B_K \le 1$.
We binned the sets which passed in the $\sa-\sb$ plane,
and thus obtained their probability
distribution. We show in Fig.~2 the resulting 68\% and
90\% probability contours in dark gray and light gray,
respectively. Since we do not know the true origin of
the CKM parameters and thus do not know the true probability
distribution from which the experimental inputs result, and since the
theoretical restrictions on $f_B$ and $B_K$ cannot
be posed statistically, we can only interpret these probability contours
as an indication of likely outcomes for $\B$-factory results based on the
SM. For example, the ``tail'' of the allowed areas which extends towards
small values of $(\sa,\sb)$ requires many of the parameters to be
stretched to their $1-\sigma$ bounds and so seems unlikely and lies
outside both probability contours.

Similarly to previous analyses (see, {\it e.g.},
\cite{nirq,agjss,schsch,ddgn,klps,kry,lmmr}),
we find that
$\sa$ can have any value in the full range from $-1$ to 1, while
$\sb$ is always positive and has a lower bound
\be
\sin2\beta\geq0.15\,.\label{abh}
\ee
Furthermore, $\sa$ is likely to be positive if the top mass is near
its present lower bound, and most importantly
{\it the favored values for $\sb$ are above 0.5}.
We also find that the bounds on the two quantities are correlated (as also
noted in \cite{ddgn}).
In particular, we note that:
\begin{itemize}
\item The magnitude of at least one of the two asymmetries is always
larger than 0.2,  and probably larger than 0.6.
\item If $\sin2\beta\leq0.4$, then $\sin2\alpha$ must be
positive---in fact, above 0.2.
\end{itemize}
Once the top mass is measured firmer predictions will of
course be possible, based on one of the graphs in Fig.~2.
\par
Various estimates may be made of the allowed ranges for the input parameters.
In particular, there is no single obvious way to
evaluate theoretical uncertainties. Furthermore, future improvement in
both experimental measurements and theoretical analyses would certainly
strengthen the constraints. Thus, it is useful to understand the
sensitivity of our analysis to the various uncertainties. To this end
we have displayed in Fig.~3 how the allowed regions of the SM depend
on the choice of input parameters, for a representative top mass of
130 GeV. For Figs.~3a, 3b and 3c we have allowed somewhat larger
ranges for $0.05\le|V_{ub}/V_{cb}|\le0.15$ and
$100\,\MeV\le f_B\le300\,\MeV$. All other ranges are kept as before. The
5 solid lines of Fig.~3a correspond, from bottom to top,
to the constraint (\ref{abbcons}) when $|V_{ub}/V_{cb}|$ increases from
$0.05$ to $0.15$. The 12 solid lines of Fig.~3b correspond,
from left to right, to the constraint (\ref{abacons}) when the values of
$f_B$ and $\tau_B|V_{cb}|^2$ decrease within their respective ranges.
The 6
solid lines of Fig.~3c correspond, from left to right, to the constraint
(\ref{abeps}) when $|V_{cb}|^2$ and $B_K$  decrease within their respective
ranges.
(Note that each solid line in these figures must meet all three constraints.
For Fig.~3b this disallows the lower end of the range for $f_B$ and
$\tau_B|V_{cb}|^2$, while for Fig.~3c it is the lower end of the range for
$|V_{cb}|^2$ and $B_K$ that is not allowed.)
One can then read off the approximate allowed
region for a more restricted choice of input
parameter ranges. For completeness we have also
plotted in Fig.~3d the allowed region obtained by accepting the range
$0.15\le|V_{ub}/V_{cb}|\le0.20$ suggested by Isgur
{\it et al.} \cite{gisw}, while keeping all other parameters as in the rest of
Fig.~3. In this case it is likely that $\sb$ is very close to unity,
or else (and this is unlikely) $\sa\sim\sb$ and they can both be as small
as roughly $0.1$ if $|V_{ub}/V_{cb}|$ and $B_K$
are as large as possible and $f_B$ is as small as possible.
\par
We next turn to the testing of various schemes for quark mass matrices.
We use the following ranges for quark masses at 1 GeV \cite{gassleut}:
\be
m_c=1.36\pm0.05\,\GeV,\ \ \ m_b=5.6\pm0.4\,\GeV,\label{galea}
\ee
and for mass ratios:
\be
{m_d\over m_s}=0.051\pm0.004,
\ \ \ {m_u\over m_c}=0.0038\pm0.0012,
\ \ \ {m_s\over m_b}=0.030\pm0.011.\label{galeb}
\ee
In the remainder of our analysis {\it we allow only} $1-\sigma$ {\it ranges
for all inputs}, since we believe that if any of these schemes need to be
stretched beyond their $1-\sigma$ predictions then their motivation is
largely lost. These $1-\sigma$ ranges should only be viewed as the {\it
favored}
values within the schemes; one should not rule out any scheme
simply on the basis that the experimental results do not quite fall
within the $1-\sigma$ predictions we obtain. In Fig.~4
we display these predictions of the four schemes for the same sample
values of $m_t$ as in Fig.~2. Only the symmetric CKM ansatz admits a
sufficiently large range of $m_t$ to be included in more than one graph.
For reference we have also indicated, in gray, the $1-\sigma$ allowed
areas of the SM.
\par
We first discuss the Fritzsch scheme \cite{frit},
\be
M_u=\left(
\begin{array}{ccc}
0&a_u&0 \\
a_u&0&b_u \\
0&b_u&c_u
\end{array}\right),\ \ \
M_d=\left(
\begin{array}{ccc}
0&a_de^{i\phi_1}&0 \\
a_de^{-i\phi_1}&0&b_de^{i\phi_2} \\
0&b_de^{-i\phi_2}&c_d \\
\end{array}\right).\label{fria}
\ee
It fits ten parameters (6 masses, 3 mixing angles
and a \cp-violating phase) with
eight parameters and therefore makes two predictions. It
is now nearly excluded \cite{hani}.
The main difficulty lies in the relation
\be
|V_{cb}|=\left|\sqrt{m_s\over m_b}-e^{-i\phi_2}\sqrt{m_c\over
m_t}\right|,\label{abi}
\ee
which can only be fulfilled if the top quark is close to the
experimental lower bound:
\be
m_t\sim90\,\GeV.\label{frib}
\ee
If the top quark is indeed this light, then the next crucial test
for the Fritzsch scheme would be its predictions for \CP asymmetries
in $\B^0$ decays.
The allowed range for $(\sa,\sb)$ is shown as the black wedge in Fig.~4a.
We find
\be
0.10\leq\sa\leq0.67;\ \ \
0.56\leq\sin2\beta\leq0.60.\label{fric}
\ee
\par
We turn next to the scheme of Giudice \cite{giud},
which requires the charged fermion
mass matrices to have the following form at the GUT scale:
\be
M_u=\left(
\begin{array}{ccc}
0&0&b \\
0&b&0 \\
b&0&a
\end{array}\right),\ \ \
M_d=\left(
\begin{array}{ccc}
0&fe^{i\phi}&0 \\
fe^{-i\phi}&d&2d \\
0&2d&c
\end{array}\right),\ \ \
M_\ell=\left(
\begin{array}{ccc}
0&f&0 \\
f&-3d&2d \\
0&2d&c
\end{array}\right).\label{giua}
\ee
This scheme fits the quark {\it and}\, lepton mass matrices with six
parameters and therefore makes seven predictions. Among them we find
\be
m_t\sim125-155\,\GeV,\ \ |V_{cb}|\sim0.048,\ \
0.07\le |V_{ub}/V_{cb}|\le0.084\
\left({130\,\GeV\over m_t}\right).\label{giub}
\ee
Note that our allowed range for $m_t$ is smaller than in ref.
\cite{giud}, due to our stronger bounds on $|V_{ub}/V_{cb}|$.
(This range is very sensitive to the bottom quark mass, and thus
could be enlarged by adopting more conservative
estimates of the uncertainty in $m_b$.)
It is not unlikely that this scheme would survive the various
measurements until a $\B$-factory starts running.
Then it allows only a narrow band in the $\sa-\sb$
plane, as shown in Fig.~4b. The overall constraint is
\be
-0.98\leq\sa\leq+1.0;\ \ \
0.2\leq\sb\leq0.7\,.\label{giuc}
\ee
However, for low $\sb$ values, there is a strong
correlation between the two asymmetries.
In particular, if $\sb\,\roughly{<}\,0.45$, then $\sa\geq0.65$.
\par
The scheme by DHR \cite{dhr}
requires that, at the GUT scale, charged fermion
mass matrices are of the following form:
\be
M_u=\left(
\begin{array}{ccc}
0&c&0 \\
c&0&b \\
0&b&a
\end{array}\right),\ \ \
M_d=\left(
\begin{array}{ccc}
0&fe^{i\phi}&0 \\
fe^{-i\phi}&e&0 \\
0&0&d
\end{array}\right),\ \ \
M_\ell=\left(
\begin{array}{ccc}
0&f&0 \\
f&-3e&0 \\
0&0&d
\end{array}\right).\label{dhra}
\ee
It has seven parameters and therefore six predictions, among which we find
({\it c.f.} \cite{bbhz})
\be
m_t\sim185\,\GeV,\ \ \
|V_{cb}|\sim0.047\,,\ \ \
|V_{ub}/V_{cb}|\sim0.065.\label{dhrb}
\ee
(Note that the latter prediction, which is at the top of the $1-\sigma$
range for this scheme, is just below our allowed range. We therefore
predict a very
narrow range of the DHR parameter $\chi$ which accounts for much of the
uncertainty in this scheme: $\chi^2\simeq4/3$.)
Thus, future measurements of $m_t$, or theoretical improvement
in determining $|V_{cb}|$ or $|V_{ub}/V_{cb}|$,
may easily exclude the DHR scheme.
If it survives these tests, then it would provide very powerful
predictions for \CP asymmetries in $\B^0$ decays.
Only a very  narrow range in the $\sa-\sb$
plane is allowed, as shown in Fig.~4d. The overall constraint is
\be
-0.58\leq\sa\leq-0.33;\ \ \
0.51\leq\sb\leq0.60\,.\label{dhrc}
\ee
Once again the values of the two asymmetries are correlated,
providing an even stronger test than implied by (\ref{dhrc}).
\par
Our last example is the symmetric ansatz \cite{brpr} for the CKM matrix,
\be
|V_{ij}|=|V_{ji}|.\label{syma}
\ee
The theoretical motivation for this ansatz is more obscure than for
the previous ans\"{a}tze. In particular, it is still to be
demonstrated that the constraints (\ref{syma}) can result from some
symmetry of the lagrangian \cite{sasa}. This ansatz leads to ({\it c.f.}
\cite{scs,rosn,kikr,tani})
\be
m_t\,\roughly{>}\,160\,\GeV,\ \ |V_{ub}/V_{cb}|
\geq|V_{cd}|/2\simeq0.11\,.\label{symb}
\ee
(This bound on $m_t$ is lower than in some previous analyses due to our higher
allowed range of $f_B$, as already remarked in \cite{harr}.)
\CP asymmetries in $\B^0$
decays would be extremely powerful in testing (\ref{syma}).
The correlation between $\sa$ and $\sb$ is
strongest here, as (\ref{syma}) leads to
\be
\rho=1/2\ \Longrightarrow\ \sin2\alpha=-2\sin2\beta\cos2\beta.
\label{symc}
\ee
For a fixed $m_t$ value, (\ref{symc}) leads to an allowed {\it curve}
in the $\sa-\sb$ plane, as shown in Figs.~4c and 4d.
For the overall bounds we find
\be
-1.0\leq\sa\leq-0.76\,;\ \ \
0.68\leq\sb\leq0.91\,.\label{symd}
\ee
The two-angle parametrization of the CKM matrix proposed by Kielanowski
\cite{kiel}
is a special case of this ansatz, in which $\eta\simeq1/(2\sqrt{3})$
(to within a few percent). Consequently ({\it c.f.} \cite{rosn,kikr})
$\sa=-\sqrt{3}/2=-\sb$,
as indicated by the small filled circle in Figs.~4c and 4d.
\par
Before concluding, let us mention a discussion of the structure
of quark mass matrices by Bjorken \cite{bjor}. His assumptions lead
to a prediction for the angle $\gamma$ of the unitarity
triangle, $\gamma\approx\pi/2$. For the asymmetries discussed
here, this implies $\sa=\sb$, which coincides with the predictions of
the superweak scenario.
A discussion of the experimental prospects of excluding
such a relation can be found in refs. \cite{wins,sowo}.
\par
To summarize, we have examined the predictions of the SM
and of various quark mass matrix schemes for
$\sa$ and $\sb$ or, equivalently, for the \CP asymmetries in
$\B\rightarrow\pi\pi$ and $\B\rightarrow\psi K_S$.
Our main results
are presented in Figs.~2 and 4. We have displayed them in the $\sa-\sb$ plane
to facilitate direct comparison with future experiments or non-standard models,
and to show the importance of the correlation between the predictions for $\sa$
and for $\sb$. (This correlation was also used in \cite{ddgn,haro}).
The predictions
are quite encouraging for experimenters:
\begin{itemize}
\item Recent improvements in theoretical calculations lead to
a lower bound on the asymmetry in $B\rightarrow\psi K_S$
of order 0.15, somewhat higher than previous analyses.
\item If the asymmetry in $B\rightarrow\psi K_S$
is close to its lower bound, than it is highly correlated
with the asymmetry in $\B\rightarrow\pi\pi$ and at least one of the
two is larger than 0.2.
\item For the asymmetries to both be small, many parameters
have to assume values close to their $1-\sigma$ bounds, which is improbable.
It is more likely that at least one of the asymmetries
is larger than 0.6.
\item Various schemes for quark mass matrices allow a much smaller
range for the two asymmetries than does the SM.
Therefore, they would be stringently tested when the asymmetries
are measured.
\end{itemize}
\par
US acknowledges partial support from the Albert Einstein
Center for Theoretical Physics
at the Weizmann Institute, and thanks the members of the particle theory
group at the Weizmann Institute for
their kind hospitality  and Lawrence
Hall for illuminating discussions.

\section*{Figure Captions}

\begin{description}

\item[Fig.~1:] The unitarity condition
$V_{ub}^*V_{ud}^{\vphantom*} + V_{cb}^*V_{cd}^{\vphantom*}
+ V_{tb}^*V_{td}^{\vphantom*}=0$ represented as a triangle in the complex
plane. The sides have been divided by $|V_{cb}V_{cd}|$ so that the vertices may
be placed at $(0,0)$, $(1,0)$ and $(\rho,\eta)$. The angles $\alpha$ and
$\beta$ are measured counterclockwise as shown.

\item[Fig.~2:] The SM predictions in the
$\sa {\rm(horizontal)}-\sb{\rm(vertical)}$ plane, for four different top quark
masses: (a) 90 GeV, (b) 130 GeV, (c) 160 GeV and (d) 185 GeV.
The regions allowed by the $1-\sigma$
ranges for all parameters described in the text are outlined by the thin black
lines. The 68\% probability contours generated as described in the text are
shown as thick dark-gray lines, while the 90\% contours are indicated by
thinner light-gray lines.

\item[Fig.~3:] The dependence of the allowed regions in the $\sa-\sb$
plane on the input parameter ranges, for a representative value of
$m_t=130\,\GeV$. The largest region allowed by all three constraints is
outlined by the dashed lines. For this figure we have allowed the wider
range $100\,\MeV\le f_B\le300\,\MeV$ but kept all other ranges as before, with
the exception of $|V_{ub}/V_{cb}|$: in 3a, b and c we allow the wider range
$0.05\le|V_{ub}/V_{cb}|\le0.15$, while in 3d we adopt the higher range
of ref.~\cite{gisw}, $0.15\le|V_{ub}/V_{cb}|\le0.20$.
The 5 solid lines in 3a correspond, from bottom to top,
to the constraint (\ref{abbcons}) when $|V_{ub}/V_{cb}|$ increases from
$0.05$ to $0.15$. The 12 solid lines of 3b correspond,
from left to right, to the constraint (\ref{abacons}) when
its right-hand side increases from 0.29 to 2.71.
The 6
solid lines of 3c correspond, from left to right, to the constraint
(\ref{abeps}) when its first bracketed expression
increases from 1.34 to 1.40 and the
second increases from 0.24 to 0.96.

\item[Fig.~4:] The $1-\sigma$ allowed regions predicted by various mass
matrix schemes in the $\sa-\sb$ plane,
for the same sample values of $m_t$ as in Fig.~2.
The allowed regions within the SM are outlined for reference in light gray.
In 4a the value of $m_t=90\,\GeV$ is consistent only with the Fritzsch ansatz,
which predicts the values within the thin black wedge. A top mass of
$m_t=130\,\GeV$ in 4b is compatible only with the scheme of Giudice, which
allows the region within the band outlined in black. A symmetric CKM matrix is
consistent with a top mass of 160 GeV (4c) and 185 GeV (4d); its predictions
lie along the short black curve, while the special case of Kielanowski is shown
as the small filled circle in each of these figures. The DHR scheme
predicts the heavy top mass $m_t=185\,\GeV$ of 4d, and allows only the tiny
region shown in black.

\end{description}

\end{document}